4th International Workshop on Plasticity, Damage and Fracture of Engineering Materials (IWPDF 2025)

# Shear band patterns by boundary integral equations

Davide Bigoni[a], Domenico Capuani[b]*

*[a]DICAM, University of Trento, via Mesiano 77, I-38123 Trento, Italy*
*[b]DA, University of Ferrara, via Quartieri 8, I-44121 Ferrara, Italy*

**Abstract**

Boundary integral equations are presented to analyze perturbations in terms of small elastic deformations superimposed upon an arbitrary, homogeneous strain. Plane strain deformations of an incompressible, prestressed, anisotropic, elastic solid are considered assuming the Biot constitutive framework. The special case of perturbations of stress/deformation incident wave fields, caused by a shear band of finite length formed inside the material at a certain stage of the deformation path, is formulated.
© 2026 The Authors. Published by ELSEVIER B.V.
This is an open access article under the CC BY-NC-ND license (https://creativecommons.org/licenses/by-nc-nd/4.0)
Peer-review under responsibility of IWPDF 2025 scientific committee members
*Keywords:* Shear bands; pre-stress; nonlinear elasticity

## 1. Introduction

Large strain effects are important in a number of engineering problems, and they can influence the behaviour of microelectromechanical systems, geological formations, biological tissues, and structural elements, such as seismic insulators and rubber bearings. High levels of pre-stress may induce shear bands in materials like metals, polymers, granular solids.

Localized deformations in the form of shear bands are known to be preferential near-failure deformation modes of ductile materials (see Rice (1973)). The development of shear localization bands has been also shown to be possible in anisotropic composite materials consisting of random distributions of aligned rigid fibres of elliptical cross section in a soft elastomeric matrix (see Avazmohammadi and Ponte Castañeda (2016)).

* Corresponding author. Tel.: +39-532-293620 ; fax: +39-532-293620.
  *E-mail address:* domenico.capuani@unife.it

Although it is expected that dynamic effects play an important role on shear band growth, most of the analyses conducted so far were limited to quasi static conditions. Propagation of shear bands was investigated for quasi-static deformation processes showing that different shear band patterns emerge as related to load conditions and material properties of samples (see He et al. (2016), Qu et al. (2015)). In dynamics, numerical simulations were restricted to the case of high strain-rate loading (see Bonnet-Labouvier et al. (2002)).

The object of the present paper is the study of the dynamic perturbations, induced by a shear band of finite length, on time-harmonic incident wave fields traveling through an infinite medium. The shear band of finite length is idealized as a discontinuity surface formed in the medium at a certain stage of a deformation path, and it is seen as a weak surface whose faces can freely slide but are constrained to remain in contact. On the basis of the infinite-body Green functions developed by the Authors for incremental elastic deformations (see Bigoni and Capuani (2002), Bigoni and Capuani (2005)), the integral representations relating incremental displacement and incremental stress at any point of the medium to the incremental displacement jump across the shear band faces, are given. Once the boundary conditions at the shear band surface are enforced, the boundary integral equation, determining the unknown jump of tangential incremental displacement across the shear band faces, can be obtained.

The integral equations are the means to analyze strain localization as a special case of instability which is induced by perturbations and is found to occur within the elliptic range.

## 2. Constitutive equations

The incremental behaviour of an infinite, incompressible, nonlinear elastic material, homogeneously deformed under plane strain condition, is considered. According to Biot (1965), the constitutive relations between the nominal stress increment $\dot{t}_{ij}$ and the gradient of incremental displacement $v_{i,j}$ (a comma denotes partial differentiation) can be expressed in the principal reference system of Cauchy stress (here denoted by axes $x_1$ and $x_2$) as follows

$$\dot{t}_{ij} = K_{ijkl}v_{l,k} + \dot{p}\,\delta_{ij} \tag{1}$$

where repeated indices are summed and range between 1 and 2, $\delta_{ij}$ is the Kronecker delta, $\dot{p}$ is the incremental hydrostatic stress and $K_{ijkl}$ are the instantaneous moduli.

These moduli possess the major symmetry $K_{ijkl} = K_{klij}$ and are functions of principal components of Cauchy stress, $\sigma_1$ and $\sigma_2$, describing the pre-stress, and of two incremental moduli $\mu$ and $\mu_*$ (which can depend arbitrarily on the current stress and strain) corresponding to shearing parallel to, and at 45° to, the principal stress axes. The non-null components are:

$$K_{1111} = \mu_* - \frac{\sigma}{2} - p\,, \quad K_{1122} = K_{2211} - \mu_*\,, \quad K_{2222} = \mu_* + \frac{\sigma}{2} - p$$

$$K_{1212} = \mu + \frac{\sigma}{2}\,, \quad K_{1221} = K_{2112} = \mu - p\,, \quad K_{2121} = \mu - \frac{\sigma}{2} \tag{2}$$

with

$$\sigma = \sigma_1 - \sigma_2\,, \quad p = (\sigma_1 + \sigma_2)/2 \tag{3}$$

Equation (1) is complemented by the incompressibility constraint for incremental displacement $v_i$

$$v_{i,i} = 0 \tag{4}$$

Constitutive Eqs. (1)-(4) describe a broad class of material behaviors, including all possible elastic incompressible materials which are isotropic in an initial state, but also materials which are orthotropic with respect to the principal stress directions.

## 3. The boundary value problem

In Figure 1, a shear band of total length $2l$ is represented together with a local reference system $(\hat{x}_1, \hat{x}_2)$ centered on each shear band, with $\hat{x}_1$-axis aligned parallel to the shear band, and rotated at an angle θ with respect to the principal reference system $(x_1, x_2)$ introduced for constitutive Eq. (1).

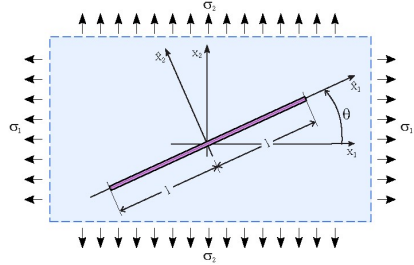

Fig. 1. Shear band of finite length ($2l$) and principal Cauchy stress components $\sigma_1$.and $\sigma_2$.

According to the model described in Giarola et al. (2018), by introducing the jump operator for a generic function $f$, smooth on two regions labeled "+" and "-", and discontinuous across the surface $S$ of the shear band, as

$$[\![f]\!] = f^+ - f^- \tag{5}$$

where $f^\pm$ denote the limits approached by function $f$ at the faces of the discontinuity surface, the boundary conditions at shear band surface $S$ can be written as

$$[\![\hat{v}_2]\!] = 0 \,, \quad [\![\hat{t}_{22}]\!] = 0 \,, \quad \hat{t}_{21} = 0 \tag{6}$$

with $\hat{v}_i, \hat{t}_{ij}$ being incremental displacement and incremental stress components in the local reference system.

Time-harmonic incident shear waves of circular frequency $\Omega$ characterized by incremental displacement field $\mathbf{v}^{inc}(\mathbf{x})$, with amplitude $A$ and phase velocity $c$, propagation direction $\mathbf{p}$ and direction of motion $\mathbf{d}$, are considered

$$\boldsymbol{v}^{inc} = A\boldsymbol{d}e^{i\frac{\Omega}{c}(\mathbf{x}\mathbf{p}-ct)} \tag{7}$$

so that the total incremental displacement field $\mathbf{v}(\mathbf{x})$ is given by the sum of the incident and of the scattered field $\mathbf{v}^{sc}(\mathbf{x})$:

$$\boldsymbol{v} = \boldsymbol{v}^{inc} + \boldsymbol{v}^{sc} \tag{8}$$

## 4. Boundary integral equations

With reference to any given point $\mathbf{y}$ outside the shear band, the scattered field can be given the following integral representation in terms of jumps of incremental displacements and incremental stress across the discontinuity surface

$$v_g^{sc}(\boldsymbol{y}) = \int_S \ t_{ij}^g(\boldsymbol{x}, \boldsymbol{y}) n_i [\![v_j]\!] dl_x \tag{9}$$

where $\mathbf{n}$ is the unit normal at every point of $S$, and $t_{ij}^g(\mathbf{x}, \mathbf{y})$ is the incremental nominal stress associated with the incremental displacement $v_j^g(\mathbf{x}, \mathbf{y})$ of the time-harmonic Green function for the infinite prestressed medium found by Bigoni and Capuani (2005).

Denoting by **s** the unit tangent vector at any point of $S$, the boundary condition $(6_1)$ leads to

$$\llbracket v_j \rrbracket = s_j \llbracket \hat{v}_1 \rrbracket \tag{10}$$

and considering that

$$\hat{t}_{21} = \dot{t}_{ij} n_i s_j \tag{11}$$

equation (9) can be rewritten as

$$v_g^{sc}(\boldsymbol{y}) = \int_{-l}^{l} \hat{t}_{21}^g(\hat{x}_1, \boldsymbol{y}) \llbracket \hat{v}_1 \rrbracket d\hat{x}_1 \tag{12}$$

The gradient of incremental displacement can be evaluated from (9) as

$$v_{g,k}^{sc}(\boldsymbol{y}) = -\int_{-l}^{l} \hat{t}_{21,k}^g(\hat{x}_1, \boldsymbol{y}) \llbracket \hat{v}_1 \rrbracket d\hat{x}_1 \tag{13}$$

and the incremental stress can be deduced from constitutive equations (1):

$$\dot{t}_{lm}^{sc} = K_{lmkg} v_{g,k}^{sc} + \dot{p}\delta_{lm} \tag{14}$$

By enforcing the boundary condition $(6_3)$ for point **y** approaching a point of $S$, implies

$$\hat{t}_{21}^{sc} = -\hat{t}_{21}^{inc} \tag{15}$$

and, using equations (13) and (14),

$$\hat{t}_{21}^{inc} = K_{lmkg} n_l s_m \int_{-l}^{l} \hat{t}_{21,k}^g(\hat{x}_1, \boldsymbol{y}) \llbracket \hat{v}_1 \rrbracket d\hat{x}_1 \tag{16}$$

Equation (16) represents the boundary integral formulation for the boundary value problem at hand. The kernel of the integral equation (20) is hypersingular of order $r^{-2}$ as $r \to 0$, $r$ being the distance between field point **x** and source point **y**. Therefore, the integral on right-hand side of (20) is specified in the finite-part Hadamard sense.

Equation (16) shows that the solution of the problem is given by a linear integral equation in the unknown scalar function $\llbracket \hat{v}_1 \rrbracket$, i.e. the jump of tangential incremental displacement across the shear band faces.

## 5. Numerical example

Since both field and source points **x** and **y** lie on the $\hat{x}_1$–axis, equation (16) can be rewritten as

$$\hat{t}_{21}^{inc}(\hat{y}) = n_l s_m K_{lmkg} \int_{-l}^{l} \hat{t}_{21,k}^g(\hat{x}, \hat{y}) \llbracket \hat{v} \rrbracket d\hat{x} \tag{17}$$

where the index '1' has been dropped, so that $\hat{x}$, $\hat{y}$, and $\llbracket \hat{v} \rrbracket$ replace respectively $\hat{x}_1$, $\hat{y}_1$ and $\llbracket \hat{v}_1 \rrbracket$.
Using equations (2), equation (17) becomes

$$\hat{t}_{21}^{inc}(\hat{y}) = -2 \sin\theta \cos\theta (2\mu_* - p) \int_{-l}^{l} \hat{t}_{21,1}^1(\hat{x}, \hat{y}) \llbracket \hat{v} \rrbracket d\hat{x}$$

$$-(\sin\theta)^2 \left[ (\mu - p) \int_{-l}^{l} \hat{t}_{21,2}^1(\hat{x}, \hat{y}) \llbracket \hat{v} \rrbracket d\hat{x} + \mu(1+k) \int_{-l}^{l} \hat{t}_{21,1}^2(\hat{x}, \hat{y}) \llbracket \hat{v} \rrbracket d\hat{x} \right]$$

$$+(cos\,\theta)^2\left[(\mu-p)\int_{-l}^{l}\hat{t}_{21,1}^2(\hat{x},\hat{y})[\![\hat{v}]\!]d\hat{x}+\mu(1-k)\int_{-l}^{l}\hat{t}_{21,2}^2(\hat{x},\hat{y})[\![\hat{v}]\!]d\hat{x}\right]$$ (18)

A low frequency approximation for $[\![\hat{v}]\!]$ can be found by adopting

$$[\![\hat{v}]\!](\hat{x})=\hat{v}_o\sqrt{1+\frac{\hat{x}}{l}}\sqrt{1-\frac{\hat{x}}{l}}$$ (19)

so that the unknown turns out to be the amplitude of the incremental displacement jump across the shear band faces.

A ductile low-hardening metal, modelled within the $J_2$-deformation theory (see Hutchinson and Neale (1979)), with the hardening exponent $N$=0.4, is considered. In the $J_2$-deformation theory, which is particularly suited to analyze the loading branch of the constitutive response of ductile metals, the prestress parameter $k=\sigma/2\mu$, and the orthotropy parameter $\xi=\mu_*/\mu$, are given by the relations

$$k=\frac{\lambda^4-1}{\lambda^4+1}\,,\quad\xi=\frac{N(\lambda^4-1)}{2(ln\,\lambda)(\lambda^4+1)}$$ (20)

where $N$ is the hardening exponent, and $\lambda$ is the logarithmic stretch representing a prestrain measure. In the case of $N$=0.4, failure of ellipticity occurs at a prestress level $k$=0.8753.

A plane incident wave field travelling parallel to shear band, is considered, with a wavelength $\lambda_1=2\pi l$, where $\lambda_1$ corresponds to the wavelength of a plane transverse wave propagating parallel to $x_1$-axis with propagation speed $c$, i.e.

$$c=\sqrt{\frac{\mu(1+k)}{\rho}}\,,\quad\lambda_1=2\pi\frac{c}{\Omega}$$ (21)

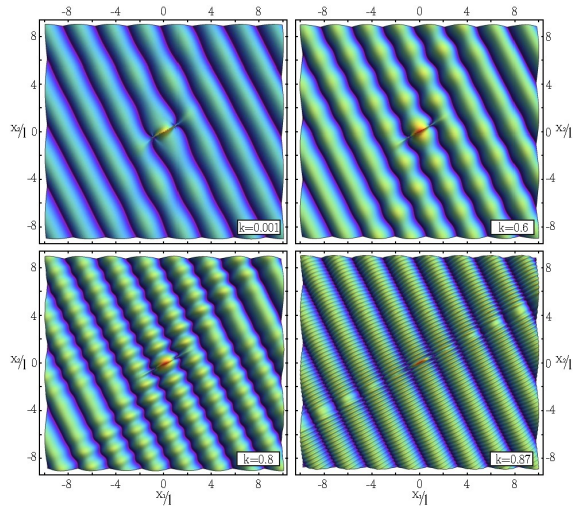

Fig. 2. Real part of total incremental deviatoric strain at different prestress levels.

The material dynamic response is shown in terms of level sets of incremental deviatoric strain. Level sets of the real part of incremental deviatoric strain are reported in Figure 2 for the total wave field. Depending on the level of prestress and anisotropy, wave patterns are shown to emerge, with focussing of signals in the direction of shear bands. It can be seen that the incident wave field is corrugated with ridges parallel to the shear band, which become closer and closer when the prestress increases.

**Acknowledgements**

Financial support from the ERC advanced grant ERC-ADG-2021-101052956-BEYOND and from the University of Ferrara (FAR) is gratefully acknowledged.

**References**

Avazmohammadi, R., Ponte Castañeda, P., 2016. Macroscopic constitutive relations for elastomers reinforced with short aligned fibers: instabilities and post-bifurcation response. Journal of the Mechanics and Physics of Solids 97, 37-67.

Bigoni, D., Capuani, D., 2002. Green's function for incremental nonlinear elasticity: shear bands and boundary integral formulation. Journal of the Mechanics and Physics of Solids 50, 471-500.

Bigoni, D., Capuani, D., 2005. Time-harmonic Green's function and boundary integral formulation for incremental nonlinear elasticity: dynamics of wave patterns and shear bands. Journal of the Mechanics and Physics of Solids 53, 1163-1187.

Biot, M.A., 1965. Mechanics of Incremental Deformations. Wiley, New York.

Bonnet-Lebouvier, A.S., Molinari, A., Lipinski, P., 2002. Analysis of the dynamic propagation of adiabatic shear bands. International Journal of Solids and Structures 39, 4249-4269.

Giarola, D., Capuani, D., Bigoni, D., 2018. The dynamics of a shear band. Journal of the Mechanics and Physics of Solids 112, 472-490.

He, J., et al., 2016. Local microstructure evolution at shear bands in metallic glasses with nanoscale phase separation. Scientific Reports 6, 25832.

Hutchinson, J.W., Neale, K.W., 1979. Finite strain J2-deformation theory. In "Proceedings of the IUTAM Symposium on Finite Elasticity". In: Carlson, D.E., Shield, R.T. (Eds.), Martinus Nijoff, The Hague, pp. 237-247.

Qu, R.T., Liu, Z.Q., Wang, G., Zhang, Z.F., 2015. Progressive shear band propagation in metallic glasses under compression. Acta Materialia 91, 19-33.

Rice, J.R., 1973. The initiation and growth of shear bands. In "Plasticity and Soil Mechancs". In: Palmer, A.C. (Ed.), Cambridge University Engineering Department, Cambridge, UK, pp. 263-274.